\documentclass{kapproc} 
\kluwerbib

\let\lcitebracket(
\let\rcitebracket)

\usepackage{graphicx}

\begin{document}

\articletitle[Spin-dependent transport in phase-separated
manganites]{Spin-dependent transport\\ in phase-separated
manganites}

\chaptitlerunninghead{Spin-dependent transport in manganites}

\author{K.\,I.\,Kugel, A.\,L.\,Rakhmanov, A.\,O.\,Sboychakov}
\affil{Institute for Theoretical and Applied Electrodynamics,
Russian Academy of Sciences\\ Izhorskaya str. 13/19, Moscow
125412, Russia} \email{kugel@orc.ru}

\author{M.\,Yu.\,Kagan, I.\,V.\,Brodsky, A.\,V.\,Klaptsov}
\affil{Kapitza Institute for Physical Problems, Russian Academy of
Sciences\\ Kosygina str. 2, Moscow, 119334 Russia}


 \begin{abstract}
Starting  from the assumption that ferromagnetically correlated
regions exist in manganites even in the absence of long-range
magnetic order, we construct a model of charge transfer due to the
spin-dependent tunnelling of charge carriers between such regions.
This model allows us to analyze the temperature and magnetic field
dependence of resistivity, magnetoresistance, and magnetic
susceptibility of phase-separated manganites in the temperature
range corresponding to non-metallic behavior. The comparison of
theoretical and experimental results reveals the main
characteristics of the phase-separated state.

\end{abstract}


\section{Introduction}
Unusual properties and the richness of the phase diagram of
manganites gave rise to a huge number of papers dealing with
different aspects of the physics of these compounds. A special
current interest to  manganites is related to the possible
existence of various inhomogeneous charge and spin states such as
lattice and magnetic polarons, droplet and stripe structures, etc.
\cite{Dagotto2001,Nagaev2001,Kagan2001}. Analogous phenomena are
well known for many strongly correlated systems where the
electron-electron interaction energy is higher than the kinetic
energy. One of the most spectacular manifestations of such a
behavior, i.e. the formation of ferromagnetic (FM) droplets
(ferrons) was predicted in Ref.~\cite{Nagaev1967} for low-doped
antiferromagnetic (AFM) semiconductors. Another example is a
formation of a string (linear trace of frustrated spins) upon the
motion of a hole in an AFM isolator \cite{Bulaevskii1968}. Both
these examples refer to the so-called electron phase separation,
when a single charge carrier changes locally its electronic
environment. In addition to this nanoscale phase separation,
manganites can also exhibit a large-scale phase separation
corresponding to the coexistence of different phases
characteristic of first-order phase transitions (e.g., the
transition between AFM and FM states). An example of this
large-scale phase separation is given by the formation of
relatively large FM droplets inside the AFM matrix. These droplets
with linear sizes of about 100-1000\,\AA\, were observed in
several experiments, in particular, by neutron diffraction methods
in Ref.~\cite{Balagurov2001}. Note also that the attraction
between one-electron ferromagnetic droplets (mediated by either
elastic or magneto-dipole interaction) can result in merging of
the ferrons and formation of intermediate to large-scale
inhomogeneities \cite{Lorenzana2001}. There exist clear
experimental indications suggesting that the phase separation is
inherent for both magnetically ordered phases and the paramagnetic
state \cite{Dagotto2001,Nagaev2001,Kagan2001,Solin2003}.
Therefore, the formation of inhomogeneous states proved to be a
typical phenomenon for manganites in different parts of their
phase diagram. Moreover, the phase separation should strongly
affect the magnetic and transport properties of manganites.

Phase separation arguments are most often used for the domain of
the existence of antiferromagnetism and especially in the vicinity
of a transition between AFM and FM states. However, as we
mentioned earlier, a manganite can be inhomogeneous even in the
paramagnetic state at temperatures exceeding the corresponding
phase transition temperature. An analysis of experimental data
reveals a substantial similarity in the high-temperature behavior
of resistivity, magnetoresistance, and magnetic susceptibility for
various manganites with different low-temperature states
\cite{Babushkina2003,Fisher,Wagner2002,Zhao2001}. In addition, the
magnetoresistance turns out to be rather large far from the FM-AFM
transition and even in the paramagnetic region. Furthermore, the
magnetic susceptibility of manganites is substantially higher than
that for typical antiferromagnets. These experimental data clearly
suggest the existence of significant FM correlations in the
high-temperature range.

Here, we start from the assumption that the ferromagnetically
correlated regions exist in manganites above the temperatures
characterizing the onset of the long-range magnetic (FM or AFM)
ordering. This assumption allows us to describe the characteristic
features of resistivity, magnetoresistance, and magnetic
susceptibility of manganites in the non-metallic state within the
framework of one model. Below, we base our discussion on the model
of conductivity of phase-separated manganites developed in
Ref.~\cite{Babushkina2003,Rakhmanov2001,Sboychakov2002,Sboychakov2003}
and use experimental data for manganites of different compositions
reported in Ref.~\cite{Babushkina2003,Fisher,Wagner2002,Zhao2001}.
Note that in this paper we do not limit ourselves by consideration
of only one-electron magnetic droplets (ferrons) but rather
generalize previously obtained results to the case of arbitrary
number of electrons in ferromagnetically correlated domains.

In Section~2, the temperature dependence of resistivity is
analyzed for the inhomogeneous state with the density of
FM-correlated regions being far from the percolation threshold. In
Sections~3 and 4, within the same assumptions, we discuss the
magnetoresistance of manganites and their magnetic susceptibility,
respectively. As a result, it is shown that the model of
inhomogeneous state provides a good description for the
high-temperature behavior of manganites.The comparison of
theoretical results and experimental data allows us to reveal the
general characteristics of ferromagnetically correlated regions.

\section{Resistivity}
\begin{figure}[ht]
    \includegraphics{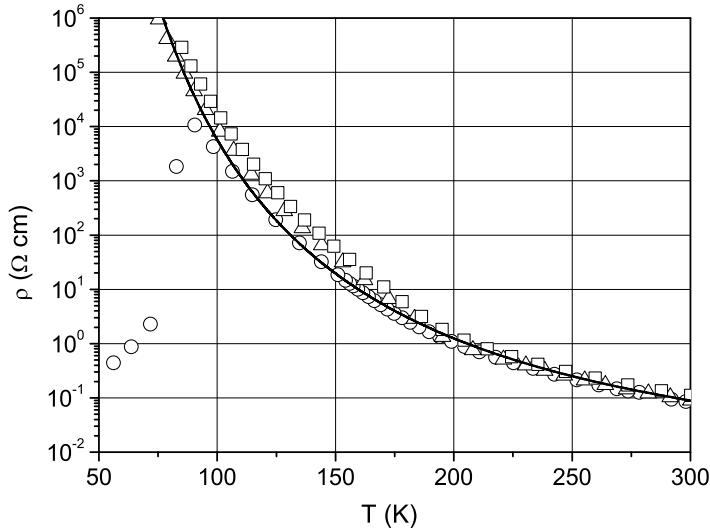}
    \caption{\label{figure1} Temperature dependence of the resistivity
    for (La$_{1-y}$Pr$_y$)$_{0.7}$Ca$_{0.3}$MnO$_3$
    samples~\cite{Babushkina2003}.
    Squares, triangles, and circles correspond to $y = 1$ (with
    $^{16}$O $\rightarrow$ $^{18}$O isotope substitution), $y = 0.75$ (with
    $^{16}$O $\rightarrow$ $^{18}$O isotope substitution), and $y = 0.75$ (with
    $^{16}$O), respectively. Solid line is the fit based on
    Eq.~(\ref{first}).}
\end{figure}
\begin{figure}[ht]
    \includegraphics{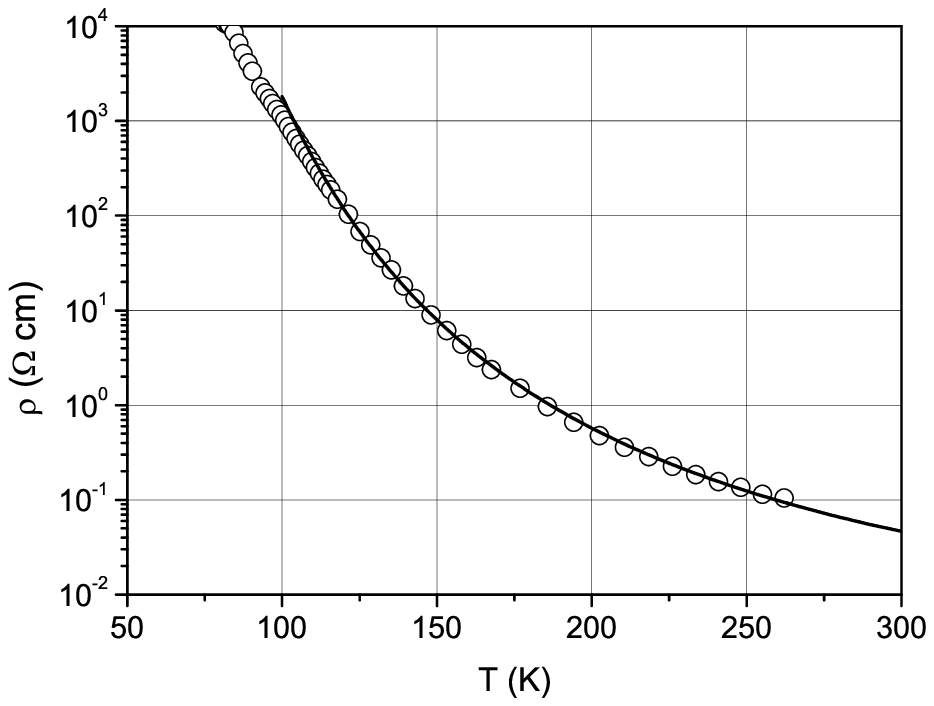}
    \caption{\label{figure2} Temperature dependence of the resistivity
    for Pr$_{0.71}$Ca$_{0.29}$MnO$_3$ sample~\cite{Fisher}: experimental
    data (circles) and theoretical curve (solid line) based on Eq.~(\ref{first}).}
\end{figure}

In the analysis of the temperature dependence of resistivity, we
will have in mind the physical picture discussed in the paper
\cite{Rakhmanov2001}. That is, we consider a non-ferromagnetic
insulating matrix with small ferromagnetic droplets embedded in
it. Charge transfer occurs via tunnelling of charge carriers from
one droplet to another. A tunnelling probability depends, strictly
speaking, upon applied magnetic field. We assume that the droplets
do not overlap and the whole system is far from the percolation
threshold. Each droplet can contain $k$ charge carriers. When a
new charge carrier tunnel to a droplet, it encounters with the
Coulomb repulsion from the carriers already residing at this
droplet. The repulsion energy $A$ is assumed to be relatively
large ($A > k_BT$). In this case, the main contribution to the
conductivity is related to the processes involving the droplets
containing $k$, $k+1$, or $k-1$ carriers. The corresponding
expression for the resistivity $\rho(T)$ has the form
\begin{equation}\label{first}
    \rho=\frac{k_BT\exp(A/2k_BT)}{128\pi e^2\omega_0 l^5kn^2},
\end{equation}
where $e$ is the charge of the electron, $\omega_0$ determines the
characteristic energy of electrons in a droplet, $l$ is the
characteristic tunnelling length, and $n$ is the concentration of
ferromagnetic droplets. Expression (\ref{first}) could be easily
derived by the method described in Ref.~\cite{Rakhmanov2001}. This
expression is a straightforward generalization of the
corresponding formula for the conductivity obtained for the case
of one-electron droplets \cite{Rakhmanov2001}. Electrical
resistivity (\ref{first}) exhibits a thermoactivation behavior
where activation energy is equal to one half of the Coulomb
repulsion energy (for details see Ref.~\cite{Rakhmanov2001}).

\begin{figure}[ht]
    \includegraphics{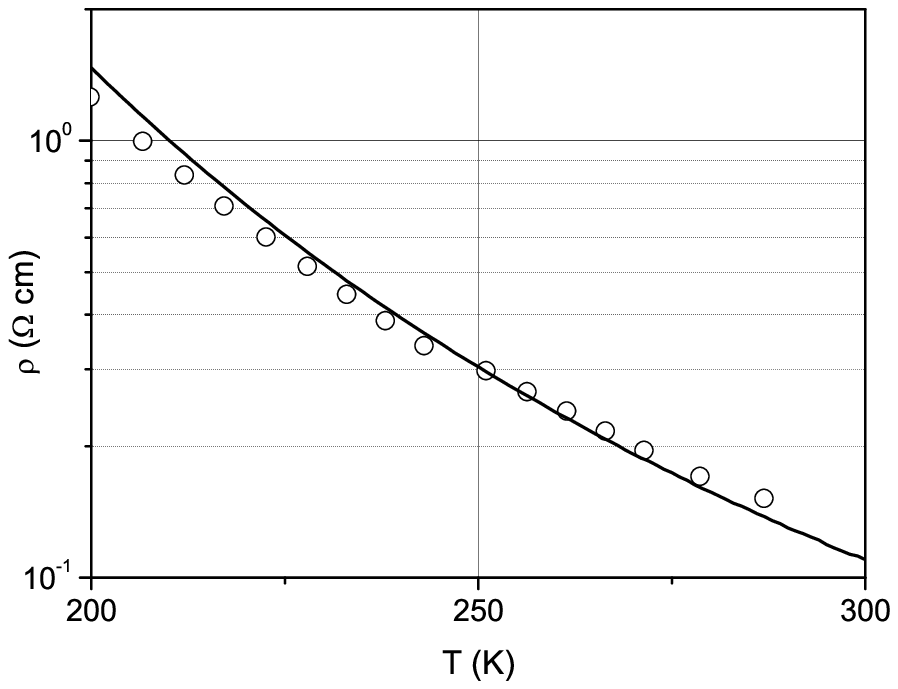}
    \caption{\label{figure3} Temperature dependence of the resistivity
    for a layered manganite (La$_{0.4}$Pr$_{0.6}$)$_{1.2}$Sr$_{1.8}$Mn$_2$O$_7$
    \cite{Wagner2002}: experimental data (circles) and theoretical
    curve (solid line) based on Eq.~(\ref{first}).}
\end{figure}
\begin{figure}[ht]
    \includegraphics{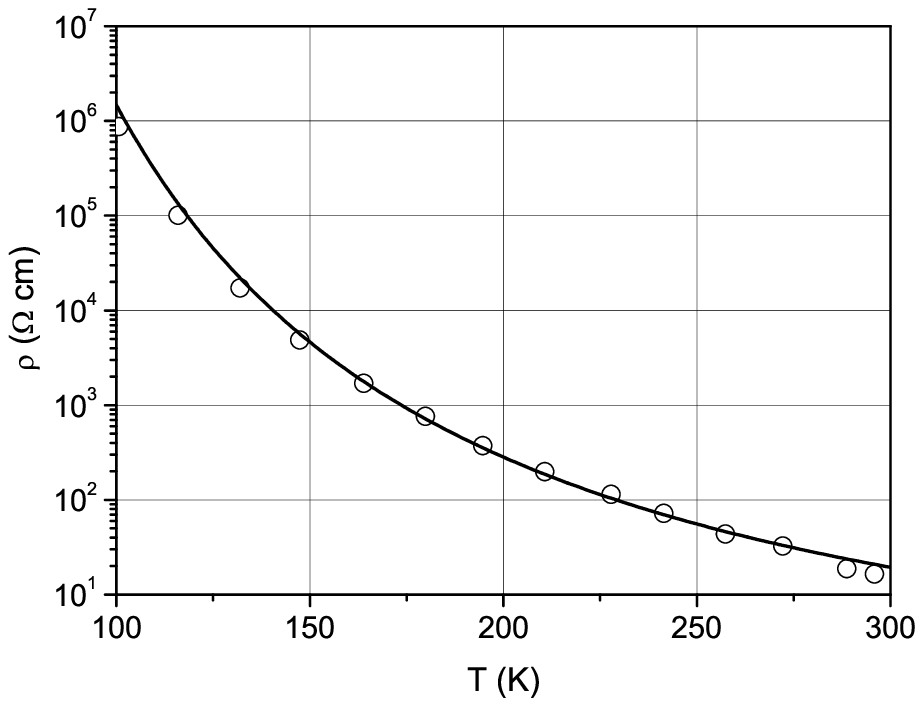}
    \caption{\label{figure4} Temperature dependence of the resistivity
    for La$_{0.8}$Mg$_{0.2}$MnO$_3$ sample \cite{Zhao2001}: experimental
    data (circles) and theoretical curve (solid line) based on
    Eq.~(\ref{first}).}
\end{figure}

Expression~(\ref{first}) provides a fairly good description for
the temperature dependence of the electrical resistivity for
various manganites. As an illustration, in
Figs.~\ref{figure1}-\ref{figure4}, we present experimental
$\rho(T)$ curves for six different materials. Experimental data
are plotted for samples reported in
Ref.~\cite{Babushkina2003,Fisher,Wagner2002,Zhao2001}. The authors
of these papers kindly provided us by the detailed numerical data
on their measurements. As it could be seen from the figures and
their captions, the examined samples differ in their chemical
composition, type of crystal structure, magnitude of electrical
resistivity (at fixed temperature, the latter varies for different
samples by more than two orders of magnitude), and also by their
low-temperature behavior (which is metallic for some samples and
insulating for the others). On the other hand, in the
high-temperature range (above the point of ferromagnetic phase
transition), $\rho(T)$ exhibits a similar behavior for all the
samples, which is well fitted by the relationship $\rho(T)\propto
T\exp(A/2k_BT)$ (solid lines in the figures).

\begin{table}[b]
    \caption{\label{Tab1}} \centering
\begin{tabular*}{\hsize}{@{\extracolsep{\fill}}lccrr}
    \sphline
    \it{Samples} & $A$, K & $\rho$(200\,K), $\Omega$$\cdot$cm
    &$l^5n^2k$, cm$^{-1}$ & \it{Data source}\cr
    \sphline
    (La$_{1-y}$Pr$_y$)$_{0.7}$Ca$_{0.3}$MnO$_3$ & 3650 & 1.25 &
    $2\cdot10^5$ & Fig. \ref{figure1} $^{a)}$\cr

    Pr$_{0.71}$Ca$_{0.29}$MnO$_3$ & 3500 & 0.57 & $3\cdot10^5$ &
    Fig. \ref{figure2} $^{b)}$\cr

    (La$_{0.4}$Pr$_{0.6}$)$_{1.2}$Sr$_{1.8}$Mn$_2$O$_7$ $^{*)}$ & 3600
    & 1.5 & $1.5\cdot10^5$ & Fig. \ref{figure3} $^{c)}$\cr

    La$_{0.8}$Mg$_{0.2}$MnO$_3$ & 3700 & 283 & $1\cdot10^3$ &
    Fig. \ref{figure4} $^{d)}$\cr
    \sphline
\end{tabular*}
\begin{tablenotes}
    $^{a)}$ \cite{Babushkina2003}

    $^{b)}$ \cite{Fisher}

    $^{c)}$ \cite{Wagner2002}

    $^{d)}$ \cite{Zhao2001}

    $^{*)}$ The chemical formula of this composition can be written as
    (La$_{0.4}$Pr$_{0.6}$)$_{2-2x}$Sr$_{1+2x}$Mn$_2$O$_7$
\end{tablenotes}
\end{table}

Based on Eq.~(\ref{first}) and experimental data, one can deduce
some quantitative characteristics of the phase-separated state. In
particular, the analysis carried out in the papers
\cite{Zhao2001,Zhao2002} demonstrated that an accurate estimate
for the value of Coulomb energy $A$ can be found by fitting
experimental data and using Eq.~(\ref{first}). The  data
represented in Fig.~\ref{figure1}-\ref{figure4} suggest that the
Coulomb barrier $A$ can be determined with an accuracy of 2-3\%
and its value lies in the narrow range from 3500 to 3700\,K (see
Table \ref{Tab1}). As it was mentioned in the papers
\cite{Zhao2001,Rakhmanov2001,Zhao2002}, the characteristic
frequency $\omega_0$ in  (\ref{first}) can also vary in a
restricted range of $10^{13}$-$10^{14}$\,Hz. This estimate might
be derived, for example, from the uncertainty principle:
$\hbar\omega_0\sim\hbar^2/2ma^2$, where $a$ is a characteristic
droplet size, and $m$ is the electron mass. Assuming $a\sim
1-2$\,nm, one obtains the latter estimate. Note also that these
values of a droplet size allow us to find an estimate for the
barrier energy $A$, which is accurate within the order of
magnitude. This energy is of the order of $e^2/\varepsilon a$, and
substituting permittivity $\varepsilon\sim10$, we get a value of
$A$ consistent with the experimental data.

It is rather difficult to estimate the tunnelling length $l$.
However, we can say that in the domain of the applicability of
relationship~(\ref{first}), length $l$ cannot be much smaller than
an interdroplet spacing \cite{Rakhmanov2001}. In another
situation, the behavior of the resistivity would be different. In
the quasiclassical approximation, the tunnelling length is of the
order of the characteristic size for the wave function provided
the barrier height is comparable with the depth of the potential
well. In our case, the size of the electron wave function is of
the order of a ferron size, while the height of the barrier
practically coincides with the depth of the potential well. The
latter naturally follows from the model of ferron formation
\cite{Nagaev2001}. Therefore, it seems reasonable to assume the
tunnelling length to be of the same order as a ferron size (few
nanometers), though, generally speaking, it can substantially
differ from $a$.

It is rather nontrivial task to estimate the concentration $n$ of
ferrons. In fact, following the papers \cite{Zhao2001,Zhao2002},
concentration $n$ could be determined by the dopant concentration
$x$ as $n\approx x/d^3$. Yet this approach would bring at least
two contradictions. First, even under the moderate concentration
of divalent element $x = 0.1-0.2$ the droplets should overlap
giving rise to the continuous metallic and ferromagnetic cluster.
However, the material could be insulating even at larger
concentrations ($x = 0.5-0.6$), at least, in a high-temperature
range. Second, as it can be seen from the experimental data, the
relation between a dopant concentration and the conductivity of
manganites is relatively complicated - for some materials changing
$x$ by a factor of two can change resistivity by two orders of
magnitude \cite{Zhao2001,Zhao2002}, for other materials $\rho(x)$
exhibits even a nonmonotonic behavior in certain concentration
ranges. Note that these discrepancies are essential not only for
our model of phase separation but also for other models dealing
with the properties of manganites (e.g., polaronic models
\cite{Ziese1998,Jakob1998}). Unfortunately, the authors of the
papers \cite{Zhao2001,Zhao2002} do not take into account these
considerations when analyzing their results from the standpoint of
the existing theories of the conductivity in manganites. The
natural conclusion is that the number of carriers, which
contribute to the charge transfer processes does not coincide with
the concentration of the divalent dopant $x$. This is particularly
obvious in the case of charge ordering when some part of the
carriers introduced by doping becomes localized and forms a
regular structure.

Therefore, using expression~(\ref{first}) and  experimental data,
we are able to obtain also the value of the combination $l^5n^2k$.
In Table~\ref{Tab1}, the values of Coulomb energy $A$, resistivity
$\rho $ at 200\,K and,  combination $l^5n^2k$ are presented. All
estimations were made based on Eq.~(\ref{first}) and the
experimental data of Fig.~\ref{figure1}-\ref{figure4}. Note that
whereas the accuracy of the estimate for $A$ is about $\pm50$\,K,
the combination $l^5n^2k$ could be estimated only by the order of
magnitude (at least, due to the uncertainty in the values of
frequency $\omega_0$).

\section{Magnetoresistance}
\begin{figure}[ht]
    \includegraphics{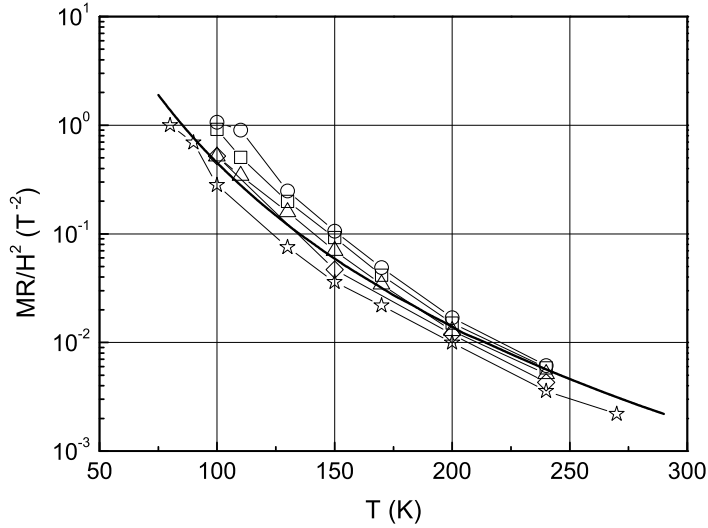}
    \caption{\label{figure5} Temperature dependence of $MR/H^2$ ratio
    for (La$_{1-y}$Pr$_y$)$_{0.7}$Ca$_{0.3}$MnO$_3$ samples
    \cite{Babushkina2003}. Squares, triangles, circles, diamonds,
    and asterisks correspond to $y = 0.75$ , $y = 0.75$ (with 30\% of
     $^{18}$O), $y = 0.75$ (with $^{16}$O $\rightarrow$ $^{18}$O isotope
     substitution), $y = 1$, and $y = 1$ (with $^{16}$O $\rightarrow$
     $^{18}$O isotope  substitution), respectively. Solid line is the
     fit based on Eq.~(\ref{MR}) ($MR \propto 1/T^5$).}
\end{figure}
\begin{figure}[ht]
    \includegraphics{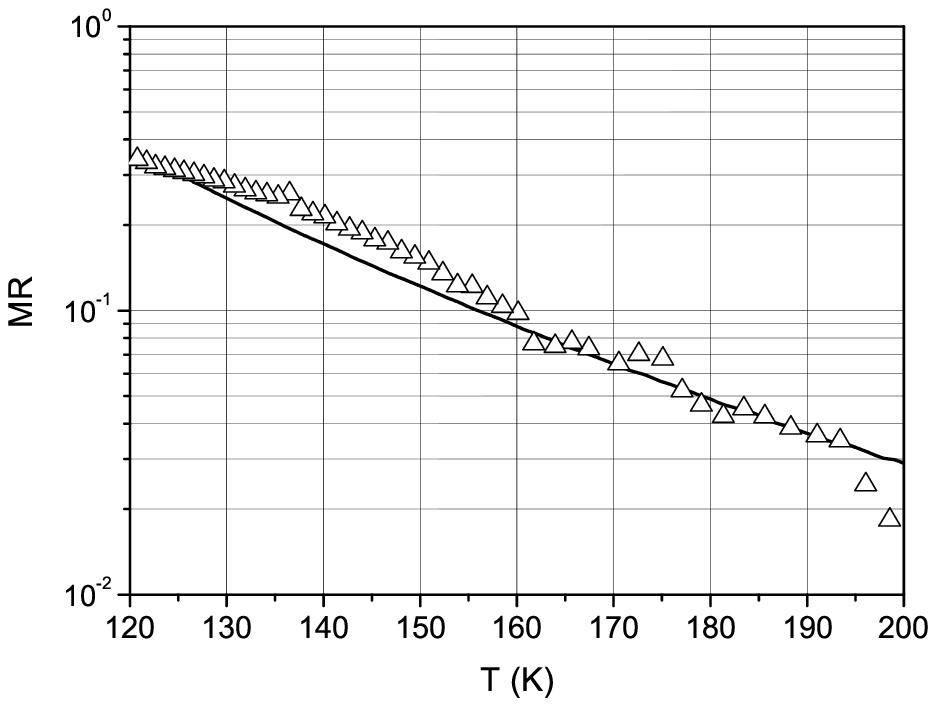}
    \caption{\label{figure6} Temperature dependence of the
    magnetoresistance for Pr$_{0.71}$Ca$_{0.29}$MnO$_3$ sample
    at $H = 2$T: experimental data (triangles) \cite{Fisher}
    and theoretical curve (solid line) based on Eq.~(\ref{MR}).}
\end{figure}

\begin{figure}[ht]
    \includegraphics{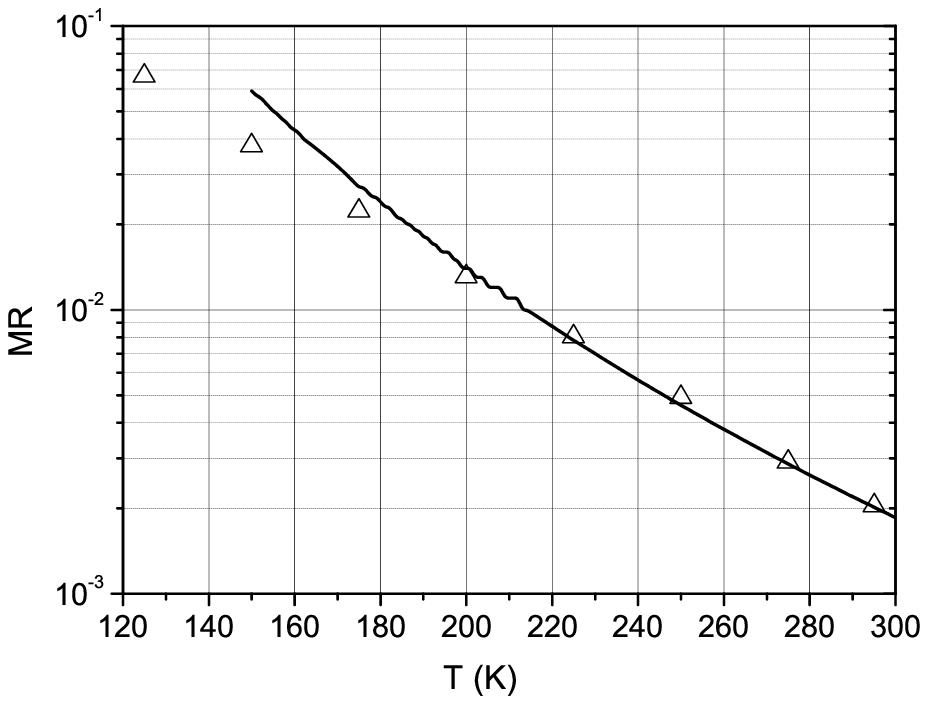}
    \caption{\label{figure7} Temperature dependence of the magnetoresistance
    for (La$_{0.4}$Pr$_{0.6}$)$_{1.2}$Sr$_{1.8}$Mn$_2$O$_7$ sample at
    $H = 1$T: experimental data (triangles) \cite{Wagner2002}
    and theoretical curve (solid line) based on Eq.~(\ref{MR}).}
\end{figure}

\begin{figure}[ht]
    \includegraphics{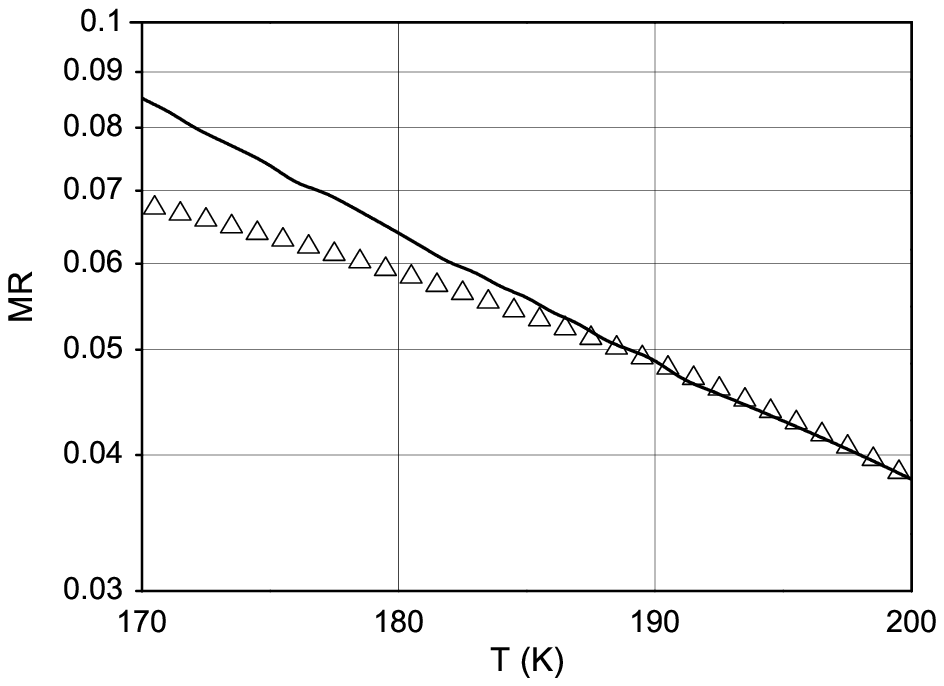}
    \caption{\label{figure8}Temperature dependence of the magnetoresistance
    for La$_{0.8}$Mg$_{0.2}$MnO$_3$ sample at
    $H = 1.5$T: experimental data (triangles) \cite{Zhao2001}
    and theoretical curve (solid line) based on Eq.~(\ref{MR}).}
\end{figure}

In the papers \cite{Babushkina2003,Sboychakov2002,Sboychakov2003},
it was demonstrated  that the model of phase separation considered
here results in a rather specific dependence of the
magnetoresistance $MR(T,H)$ on temperature and magnetic field. At
relatively high temperatures and not very strong magnetic fields,
the expression for the magnetoresistance reads

\begin{equation}\label{MR}
    MR\approx5\cdot10^{-3}\frac{\mu_B^3S^5N_{ef}^3Z^2g^3J^2H_a}{(k_BT)^5}H^2,
\end{equation}
where $\mu_B$ is the Bohr magneton, $S$ is the average spin of a
manganese ion, $N_{ef}$  is the number of  manganese atoms in a
droplet, $Z$ is the number of nearest neighbors of a manganese
ion, $g$ is the Land\'{e} factor, $J$ is the exchange integral of
the ferromagnetic interaction, and $H_a$ is the effective field of
magnetic anisotropy of a droplet. The $MR\propto H^2/T^5$
dependence was observed in the experiments for a number of
manganites in the region of their non-metallic behavior
\cite{Babushkina2003,Fisher}. The same high-temperature behavior
of the magnetoresistance can be obtained by processing the
experimental data reported in Ref.~\cite{Wagner2002,Zhao2001} (see
Figs.~\ref{figure5}-\ref{figure8}).
\begin{table}[b]
    \caption{\label{Tab2}} \centering
\begin{tabular*}{\hsize}{@{\extracolsep{\fill}}lclrr}
    \sphline
    \it{Samples} & $N_{ef}$ & $x$ & $k$ & \it{Data source}\cr
    \sphline
    (La$_{1-y}$Pr$_y$)$_{0.7}$Ca$_{0.3}$MnO$_3$ & 250 & 0.3 &
    75 & Fig. \ref{figure5} $^{a)}$\cr

    Pr$_{0.71}$Ca$_{0.29}$MnO$_3$ & 200 & 0.29 & 58 &
    Fig. \ref{figure6} $^{b)}$\cr

    (La$_{0.4}$Pr$_{0.6}$)$_{1.2}$Sr$_{1.8}$Mn$_2$O$_7$ $^{*)}$  & 250
    & 0.4 & 100 & Fig. \ref{figure7} $^{c)}$\cr

    La$_{0.8}$Mg$_{0.2}$MnO$_3$ & 265 & 0.2 & 53 &
    Fig. \ref{figure8} $^{d)}$\cr
    \sphline
\end{tabular*}
\begin{tablenotes}
    $^{a)}$ \cite{Babushkina2003}

    $^{b)}$ \cite{Fisher}

    $^{c)}$ \cite{Wagner2002}

    $^{d)}$  \cite{Zhao2001}

    $^{*)}$ The chemical formula of this composition can be written
    as (La$_{0.4}$Pr$_{0.6}$)$_{2-2x}$Sr$_{1+2x}$Mn$_2$O$_7$
\end{tablenotes}
\end{table}

The value of $S$ depends on the relative content of a trivalent
and a tetravalent manganese ions and ranges from $3/2$ to $2$.
Below it is assumed that $S=2$ for all the estimations. Parameter
$Z$ is, in fact, the number of manganese ions interacting with a
conduction electron placed in a droplet. It is reasonable to
assume that $Z$ is of the order of the number of nearest-neighbor
sites around a manganese ion, i.e. $Z\approx6$. The Land\'{e}
factor $g$ is determined from the experimental data. For
manganese, $g$ is usually assumed to be close to its spin value 2.
The exchange integral $J$ characterizes the magnetic interaction
between a conduction electron and the molecular field generated by
ferromagnetically correlated spins in a droplet. It is this
molecular field that produces a ferromagnetic state at low
temperatures. Therefore, we can use a well-known relationship
$S(S+1)ZJ/3 = k_BT_C$ of the molecular field theory to evaluate
the exchange integral (here $T_C$ is the Curie temperature). The
value of $T_C$ is determined from the experiment (based on neutron
diffraction or magnetization measurements). For example, in
La-Pr-Ca manganites, it is about $100-120$\,K
\cite{Balagurov2001}.

The magnetic anisotropy of manganites related to crystal structure
of these compounds is usually not too high. This implies that the
main contribution to the effective field of a magnetic anisotropy
$H_a$ stems from the shape anisotropy of a droplet and can be
evaluated as $H_a=\pi(1-3\tilde{N})M_s$, where $\tilde{N}$ is the
demagnetization factor of the droplet (along the main axis), $M_s$
is the magnetic moment per unit volume of the droplet. Below we
assume a droplet to be sufficiently elongated ($\tilde{N}\ll1$)
and $M_s=Sg\mu_B/d^3$. Then $H_a\approx2$\, kOe.

The value of $N_{ef}$ is determined by the size of a droplet and
it could be found from the neutron diffraction experiments.
However, we are unaware of such measurements performed for the
systems under discussion in a wide temperature range. Therefore,
$N_{ef}$ is treated here as a fitting parameter. Hence, using
Eq.~(\ref{MR}) and the above estimates, we can determine the value
of $N_{ef}$ from the experimental data on the magnetoresistance
(in the range of parameters corresponding to $MR\propto H^2/T^5$).
The results are summarized in Table \ref{Tab2}. In Figs.
\ref{figure5}-\ref{figure8}, solid curves correspond to the
fitting procedure based on Eq.~(\ref{MR}). The value of $T_C$ was
chosen to be equal to 120\, K.

As a result, the size of the ferromagnetically correlated regions
turns out to be nearly the same at temperatures about 200-300\,K
for all compositions under discussion. The volume of these regions
is approximately equal to that of a ball with 7-8 lattice
constants in diameter. It is natural to assume that within a
droplet the number of charge carriers contributing to tunnelling
processes equals to the number of dopant atoms. Hence, we can
write that $k = N_{ef}x$, where $x$ is the atomic percentage of
dopants. The values of $x$ and $k$ are presented in
Table~\ref{Tab2}.

\section{Magnetic susceptibility}

\begin{figure}[ht]
    \includegraphics{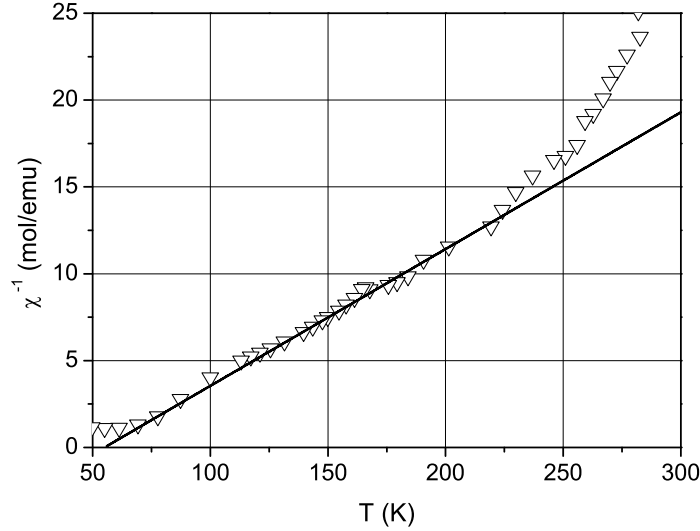}
    \caption{\label{figure9} Temperature dependence of the inverse
    magnetic susceptibility for
    La$_{1-y}$Pr$_y$)$_{0.7}$Ca$_{0.3}$MnO$_3$ sample at $y = 1$:
    experimental data (triangles) \cite{Babushkina2003}
    and theoretical curve (solid line) based on Eq.~(\ref{chi}).
    For the other samples of this group, the behavior of $\chi(T)$
    at high temperatures is rather similar to that illustrated
    in this figure (see Ref.~\cite{Babushkina2003}).}
\end{figure}

\begin{figure}[ht]
    \includegraphics{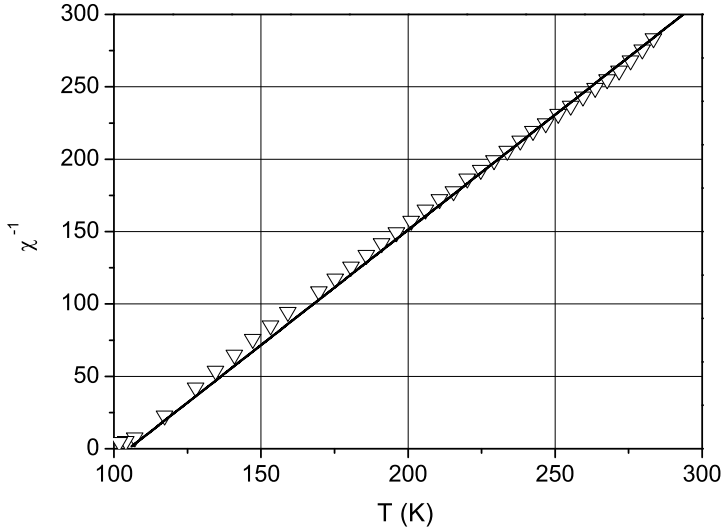}
    \caption{\label{figure10} Temperature dependence of the inverse
    magnetic susceptibility for
    Pr$_{0.71}$Ca$_{0.29}$MnO$_3$ sample:
    experimental data (triangles) \cite{Fisher} and theoretical
    curve (solid line) based on Eq.~(\ref{chi}). The sample was porous,
    its density was assumed to differ by a factor of 0.7 from the
    theoretical value.}
\end{figure}

\begin{figure}[ht]
    \includegraphics{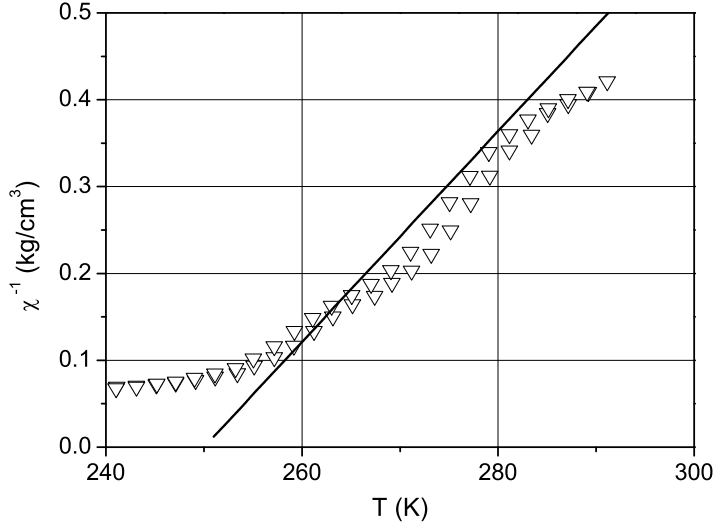}
    \caption{\label{figure11} Temperature dependence of the inverse
    magnetic susceptibility for the sample of
    (La$_{0.4}$Pr$_{0.6}$)$_{1.2}$Sr$_{1.8}$Mn$_2$O$_7$layered
    manganite: experimental data (triangles) \cite{Wagner2002}
    and theoretical curve (solid line) based on Eq.~(\ref{chi}).}
\end{figure}

\begin{figure}[ht]
    \includegraphics{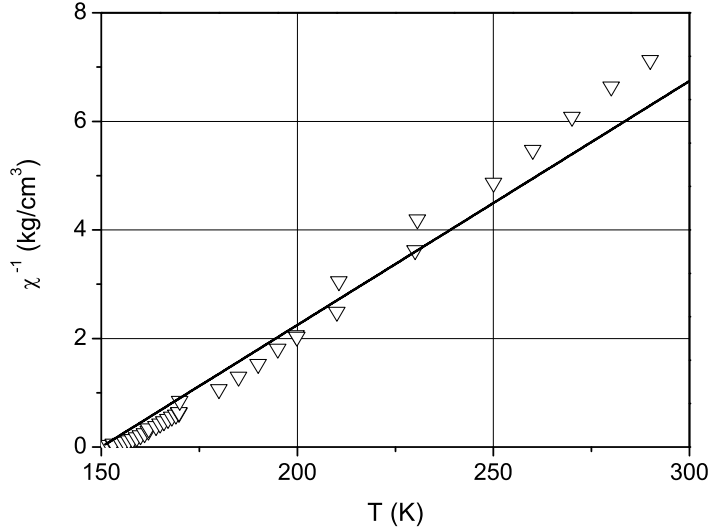}
    \caption{\label{figure12} Temperature dependence of the inverse
    magnetic susceptibility for La$_{0.8}$Mg$_{0.2}$MnO$_3$ sample:
    experimental data (triangles) \cite{Zhao2001}
    and theoretical curve (solid line) based on Eq.~(\ref{chi}).}
\end{figure}

\begin{table}[b]
    \caption{\label{Tab3}} \centering
\begin{tabular*}{\hsize}{@{\extracolsep{\fill}}lrrlcc}
    \sphline
    \it{Samples} & $\Theta$, K & $n$, cm$^{-3}$ & $p$ & $l$,
    \AA & \it{Data source}\cr
    \sphline
    (La$_{1-y}$Pr$_y$)$_{0.7}$Ca$_{0.3}$MnO$_3$ & 55 &
    $1.8\cdot10^{18}$ & 0.03 & 24 & Fig. \ref{figure9} $^{a)}$\cr

    Pr$_{0.71}$Ca$_{0.29}$MnO$_3$ & 105 &
    $6.0\cdot10^{18}$ & 0.07 & 17 & Fig. \ref{figure10} $^{b)}$\cr

    (La$_{0.4}$Pr$_{0.6}$)$_{1.2}$Sr$_{1.8}$Mn$_2$O$_7$ $^{*)}$ &  255
    & $2.5\cdot10^{18}$ & 0.04 & 19 & Fig. \ref{figure11} $^{c)}$\cr

    La$_{0.8}$Mg$_{0.2}$MnO$_3$ & 150 &
    $0.6\cdot10^{18}$ & 0.01 & 14 & Fig. \ref{figure12} $^{d)}$\cr
    \sphline
\end{tabular*}
\begin{tablenotes}
    $^{a)}$ \cite{Babushkina2003}

    $^{b)}$ \cite{Fisher}

    $^{c)}$ \cite{Wagner2002}

    $^{d)}$ \cite{Zhao2001}

    $^{*)}$ The chemical formula of this composition can be written
    as (La$_{0.4}$Pr$_{0.6}$)$_{2-2x}$Sr$_{1+2x}$Mn$_2$O$_7$
\end{tablenotes}
\end{table}

The concentration of droplets can be evaluated based on the
magnetic susceptibility data, if we assume that the dominant
contribution to the susceptibility comes from the
ferromagnetically correlated regions. At high temperatures
($k_{B}T\gg \mu_{B}gSN_{ef}H, \mu_{B}gSN_{ef}H_{a}$),
susceptibility $\chi(T)$ can be written as
\begin{equation}\label{chi}
    \chi(T)=\frac{n(\mu_BgSN_{ef})^2}{3k_B(T-\Theta)},
\end{equation}
where $\Theta$ is the Curie-Weiss constant. The results of the
processing of the experimental data are presented in
Table~\ref{Tab3}. In Figs.~\ref{figure9}-\ref{figure12}, the solid
curves correspond to the fitting procedure based on
Eq.~(\ref{chi}). Using these results, we can also estimate the
concentration of ferromagnetic phase as $p = nN_{ef}d^3$. For all
the samples, the value of the lattice constant $d$ was taken to be
equal to 3.9\,\AA. Based on the data of
Tables~\ref{Tab1}-\ref{Tab3}, it is also possible to find an
estimate for the tunnelling length~$l$.

\section{Discussion}

To sum up, the analysis performed in the previous sections
demonstrates that a simple model of the electron tunnelling
between the ferromagnetically correlated regions (FM droplets)
provides a possibility to describe the conductivity and the
magnetoresistance data for a wide class of manganites. The
comparison of the theoretical predictions with the experimental
data on the temperature dependence of the resistivity,
magnetoresistance, and magnetic susceptibility enables us to
reveal various characteristics of the phase-separated state such
as the size of FM droplets, their density, the number of electrons
in a droplet and also to estimate the characteristic tunnelling
length of the charge carriers. The determined values of parameters
appear to be rather reasonable. Indeed, the characteristic
tunnelling length turns out to be of the order of FM droplet size,
the concentration of the ferromagnetic phase in the
high-temperature range is substantially smaller than the
percolation threshold and varies from  about 1\,\% to 7\,\%.

Note also that the droplets contain 50-100 charge carriers,
whereas parameter $A$ deduced from the experimental data is equal
by the order of magnitude to the energy of Coulomb repulsion in a
metallic ball of $(7\div8)\,d$ in diameter. The obtained numerical
values for the droplet parameters (characteristic tunnelling
barrier, size, and tunnelling length) are close for manganites
with drastically different transport properties.

The large magnitude of the $1/f$ noise in the temperature range
corresponding to the insulating state is another characteristic
feature of the phase-separated manganites
\cite{Podzorov2000,Podzorov2001}. In the framework of the model of
phase separation discussed here, the following expression for the
Hooge constant was derived in the papers
\cite{Rakhmanov2001,Sboychakov2002}
\begin{equation}\label{noise}
    \alpha_H=\frac{\langle\delta
    U^2\rangle_{\omega}V_s\omega}{U^2_{DC}}=2\pi^2l^3\ln^2
    \left(\frac{\tilde{\omega}_0}{\omega}\right),
\end{equation}
where $\langle\delta U^2\rangle_{\omega}$ is the spectral density
of the voltage fluctuations, $V_s$ is the volume of a sample,
$U_{DC}$ is the applied voltage, and
$\tilde{\omega}_0=\omega_0\exp(A/2k_BT)$. Substituting to
Eq.~(\ref{noise}) the estimated values of the parameters presented
in the tables and in the text, we get $\alpha_H\approx
10^{-16}$\,cm$^3$ at temperatures 100-200\,K and frequencies
1-1000\,s$^{-1}$. This value of $\alpha_H$ is by 3-5 orders of
magnitude higher than the corresponding values for semiconductors.

Thus, we have a rather consistent scheme describing the transport
properties of manganites under condition that the
ferromagnetically correlated regions do not form a percolation
cluster. Moreover, the presented approach proves to be valid for a
fairly wide range of the dopant concentrations. However, as it was
mentioned above, the relation between the concentration of
ferromagnetic droplets and the doping level is far from being well
understood. If the picture of the phase separation is believed to
be applicable, it becomes obvious that not all electrons or holes
introduced by doping participate in the transport processes. Below
we try to present some qualitative arguments illustrating the
possible difference in the effective concentration of charge
carriers below and above the transition from paramagnetic to
magnetically ordered state.

In the phase diagram of a typical manganite, one would have the
AFM state with FM-phase inclusions in the low-temperature range
and at a low doping level. The transition from AFM to FM phase
occurs upon doping. At high temperatures, manganites are in the
paramagnetic (PM) state. When the temperature decreases, we
observe the transition from PM to AFM or FM state depending on the
doping level.

Let us consider the behavior of such a system in the vicinity of a
triple point. In the AFM phase, radius $R$ of a region which one
electron converts into FM state can be estimated as $R=d\left(\pi
t/4J_{ff}S^2Z\right)^{1/5}$ \cite{Kagan2001}, where $J_{ff}$ is an
AFM interaction constant. For high-temperature PM phase, a radius
$R_T$ of a region that one electron converts into FM state
corresponds to the size of the so-called temperature ferron and
equals to $R_T=d\left(\pi t/4k_BT\ln(2S+1)\right)^{1/5}$
\cite{Kagan2001}. The critical concentration $x_c\approx0.15$ of
the overlapping of low-temperature ferrons can be derived from the
estimate $x_c\approx 3/4\pi\cdot(d/R)^3$, while for the
high-temperature ferrons it follows from the estimate
$\delta_c\approx 3/4\pi\cdot(d/R_T)^3$. Substituting the
expressions for the radii of the high- and the low-temperature
ferrons to the ratio $x_c/\delta_c$, we obtain the following
estimate for this ratio in the vicinity of the triple point
corresponding to the coexistence of FM, AFM, and PM phases:
\begin{equation}
    \frac{x_c}{\delta_c}\sim\left[\frac{T\ln(2S+1)}{zJ_{ff}S^2}\right]^{3/5}\sim
    \left[\frac{T_C\ln(2S+1)}{T_N}\right]^{3/5},
\end{equation}
where $T_C$ and $T_N$ are the Curie and the Neel temperatures,
respectively. For the manganites under discussion, we have
$T_C\sim T_N\sim$ 120-150\,K and $\ln(2S+1)\sim1.6$ for $S=2$,
hence $\delta_c\leq x_c$. The sign of this inequality is in
agreement with experimental data which imply $\delta\sim 1-7$\,\%.
Thus, we do not have a clear explanation of the charge disbalance
in paramagnetic region in spite of the fact that the trend is
correctly caught by our simple estimates. Probably, at $x>x_c$ (in
real experiments the concentration $x$ can be as high as 50\,\%),
the residual charge is localized in the paramagnetic matrix
outside the temperature ferrons. The detailed study of this
problem will be presented elsewhere.

\begin{acknowledgments}
    The authors are grateful to V.A. Aksenov, N.A. Babushkina,
    S.W. Cheong, I. Gordon, L.M. Fisher, D.I. Khomskii, F.V. Kusmartsev,
    V.V. Moshchalkov, A.N. Taldenkov, I.F. Voloshin, G. Williams and
    X.Z. Zhou for useful discussions and provided experimental data.
    This work was supported by the Russian Foundation for Basic
    Research (Grants Nos.~02-02-16708, 03-02-06320, and NSh-1694.2003.2),
    INTAS (Grant No.~01-2008), and CRDF (Grant No.~RP2-2355-MO-02).
\end{acknowledgments}

\bibliographystyle{apalike}
\chapbblname{Kugel_Baku} \chapbibliography{Kugel_Baku}
\end{document}